\documentclass[12pt]{iopart}

\input{psfig.tex}
\newcommand{\be}{\begin{equation}}
\newcommand{\ee}{\end{equation}}
\newcommand{\bea}{\begin{eqnarray}}
\newcommand{\eea}{\end{eqnarray}}

\begin{document}

\title{Bouncing Universes with Varying Constants}
\author{John D. Barrow$^1$, Dagny Kimberly$^2$ and Jo\~ao Magueijo$^2$ }
\date{\today}

\address{$^1$DAMTP, Centre for Mathematical Sciences,
Cambridge University, Wilberforce Road,
Cambridge CB3 0WA\\
$^2$Theoretical Physics, The Blackett Laboratory,
Imperial College, Prince Consort Road, London, SW7 2BZ, U.K.}

\begin{abstract}
We investigate the behaviour of exact closed bouncing Friedmann universes in
theories with varying constants. We show that the simplest BSBM
varying-alpha theory leads to a bouncing universe. The value of alpha
increases monotonically, remaining approximately constant during most of
each cycle, but increasing significantly around each bounce. When
dissipation is introduced we show that in 
each new cycle the universe expands for
longer and to a larger size. We find a similar effect for closed bouncing
universes in Brans-Dicke theory, where $G$ also varies monotonically in time
from cycle to cycle. Similar behaviour occurs also in varying speed of light
theories.
\end{abstract}


\maketitle



The "bouncing universe"~is a modern cosmological reincarnation of an ancient
fascination with the cyclic patterns of nature and the myth of the "eternal
return" \cite{return,BT,tip}. Gravitation theories like general relativity
allow us to make precise models of this popular conception of a
\textquotedblleft phoenix\textquotedblright\ cosmology, in which a universe
periodically collapses to a Big Crunch, only to rebound into a new state of
expansion, as if emerging from a unique Big Bang \cite{tolman}. Multiple
bounces are possible but each cycle lasts longer and expands to a larger
maximum size than the previous one, a consequence of a simple application of
the second law of thermodynamics~\cite{tolman,tolman1,tolman2,zeld}, unless
there is a finite positive cosmological constant, in which case the
oscillations must eventually cease \cite{bd} and are replaced by eternal de
Sitter expansion. A sequence of many oscillations will drive the bouncing
closed universe closer and closer to flatness.

\textquotedblleft Quantum gravity\textquotedblright\ effects are invariably
invoked to justify the bounce; possible detailed calculations, however, have
only recently emerged. In loop quantum gravity, the semi-classical Friedmann
equations receive corrections that produce a bounce~\cite{singh,lid}. The
ekpyrotic model of the universe, inspired by string/M-theory, is another
possible realization of phoenix cosmology~\cite{ekpy}. It is also possible
that ghost fields -- fields endowed with negative energy -- are capable of
producing a \textit{classical} bounce (this idea has been often
rediscovered; see~\cite{gary} for a good review). Classical bounces produced
by conventional scalar fields with potentials which only violate the strong
energy condition are difficult to produce in universes that grow large
enough to be realistic: typically the probability of bounce is of order the
ratio of the minimum to the maximum expansion size \cite{star,BM}.

It has been speculated that whatever causes a collapsing universe to bounce
can reprocesses some aspects of physics, either randomly \cite{mtw}, or
systematically \cite{smol}, by changing the particle spectrum or resetting
the dimensionless \textquotedblleft constants\textquotedblright\ of Nature.
Both of these options are severely constrained by anthropic requirements but
it is interesting to ask whether there are monotonic or asymptotic trends in
the values of some quantities, as seems to be the case for the degree of
flatness of the universe, over many bounces. This matter is clearly of great
importance in the context of varying-constant theories~\cite{uzan,vslreview}%
. Here, quantities which are traditionally constants become space-time
variables and if \ singularities are avoided in the bounce then their
evolution from cycle to cycle is predictable by the field equations rather
than the outcome of effectively random reprocessing. The values of any
dimensionless 'constants' of Nature could evolve towards asymptotic
attractors if they are allowed to be variables in a self-consistent theory.
Studies of these theories are also important in assessing the stability and
level of fluctuations in a bouncing universe. It has been suggested that
thermal fluctuations in bouncing models could be the origin of the cosmic
structure~\cite{pogo,vslreview}; this would provide a distinct alternative
to an origin from vacuum quantum fluctuations in a de Sitter phase of
cosmological expansion.

Although these claims are intriguing, it is difficult to evaluate them in
the absence of a concrete model for the bounce, which is usually viewed as a
black box \ from which anything can emerge \cite{mtw}. In this paper we
examine some exactly soluble examples. First, we consider a simple bouncing
cosmology following from the simplest BSBM varying-constant theory~we
developed \cite{bsm} from Bekenstein's varying-alpha model~\cite{bek}. We
derive the result that for suitable couplings (indeed those favoured by
observations~\cite{murphy,webb}; see however~\cite{vlt}) the theory leads to
a bouncing universe. We derive similar behaviour in the Brans-Dicke theory
of a varying gravitational constant $G$~\cite{bdicke}, and in a class of
varying speed of light theories~\cite{covvsl,moffat93,am,ba,rev}. In each
case we are able to find exact solutions in the absence of dissipation and
compute the evolutionary trends in the fine structure constant and $G$ from
cycle to cycle when dissipation occurs in accord with the Second Law. We
find the interesting result that the varying constants in these theories
change monotonically from cycle to cycle when the scale factor oscillates:
the scalar fields determining the constants in each cycle do not oscillate.

Crucial to our models is the idea that cosmological fields may have a
negative energy. Such fields are called ghosts~\cite{gary}, and are far from
new, having found widespread application in the study of steady-state
cosmology~\cite{stst}, phantom dark matter~\cite{ph,phbounce}, and $\kappa $%
-essence~\cite{kappa}. Bekenstein's theory~\cite{bek} is rooted in the use
of a real scalar dielectric field $\psi $, representing the allowed
variation of the electron charge according to $e=e_{0}e^{\psi }$, where $%
e_{0}$ is the present-day value of the electron charge, and so the fine
structure 'constant' evolves with respect to its present-day value, $\alpha
_{0},$ as $\alpha =\alpha _{0}e^{2\psi }$. By redefining the electromagnetic
gauge field~\cite{bsm}, $A_{\mu }$, as $a_{\mu }=e^{\psi }A_{\mu }$, and the
electromagnetic field tensor, $F_{\mu \nu }$, as $f_{\mu \nu }=e^{\psi
}F_{\mu \nu }$, the action may be written as 
\begin{equation}
S=\int d^{4}x\sqrt{-g}\left( \mathcal{L}_{g}+\mathcal{L}_{mat}+\mathcal{L}%
_{\psi }+\mathcal{L}_{em}e^{-2\psi }\right) ,  \label{Sbsm}
\end{equation}%
where $\mathcal{L}_{\psi }=-{\frac{\omega }{2}}\partial _{\mu }\psi \partial
^{\mu }\psi $, $\mathcal{L}_{em}=-\frac{1}{4}f_{\mu \nu }f^{\mu \nu }$, and $%
\mathcal{L}_{mat}$ does not depend on $\psi $. The gravitational Lagrangian
is the usual $\mathcal{L}_{g}=\frac{1}{16\pi G}R$, with $R$ being the
curvature scalar. If the scalar coupling satisfies $\omega <0$, then $\psi $
has a negative kinetic energy term, and is thus a ghost field.

This theory can fit the observational constraints on varying alpha reported
by~\cite{murphy,webb} as well as others~\cite{bsm,wep}. However for this to
be possible in the ghost-free case with $\omega >0$ one has to choose a very
special type of dark matter, in which the magnetostatic energy $B^{2}$
dominates over the electrostatic energy $E^{2}$. Even though a dark-matter
candidate was found satisfying this condition (superconducting cosmic
strings), most types of matter, including baryonic matter, are $E^{2}$
dominated, for which observational data implies $\omega <0$, so that $\psi $
is a ghost field.

Ghosts have been criticized on a variety of grounds. Classically they are a
source of instabilities if coupled to other forms of matter, since they will
try to off-load an infinite amount of positive energy into them. This is not
necessarily cataclysmic if the rate of these processes is sufficiently slow.
For instance, in steady-state cosmology to negative probabilities. 
At the quantum level, ghosts lead to
negative norm states and so negative probabilities. The
quantum instabilities are also much more severe and are present even without
direct coupling to matter, for example in runaway particle production via
the graviton vertex. Hence, at the quantum level ghosts \textit{are}
pathological.

However, we know that quantisation of the field $\psi $ is pathological even
for $\omega >0$. For one thing, the theory is non-renormalisable. The
attitude to $\psi $ should therefore be similar to that with regards to
gravity: \textquotedblleft don't quantise\textquotedblright . General
relativity is also, at face value, non-renormalisable: for instance, the
quantum corrections to the relativistic precession of the perihelion of
Mercury are infinite. This doesn't stop the classical theory from being very
successful. It may be that the quantisation of ghosts is simply more subtle;
ghosts have been found as type $II^{\ast }$ string theories~\cite{hull}.

Non-relativistic matter and the cosmological constant may be neglected near
a bounce, so let us consider a Friedmann-Robertson-Walker (FRW) universe
filled with radiation and a dielectric field, $\psi $. The cosmological
equations for BSBM varying-$\alpha $ theory (\ref{Sbsm}) are 
\begin{eqnarray}
H^{2} &=&\frac{1}{3}\left( \rho _{r}e^{-2\psi }+\rho _{\psi }\right) -{\frac{%
K}{a^{2}}},  \label{fried1} \\
{\frac{\ddot{a}}{a}} &=&-{\frac{1}{6}}\left( 2\rho _{r}e^{-2\psi }+4\rho
_{\psi }\right) ,  \label{fried2}
\end{eqnarray}%
where we have set $8\pi G=1$, $H\equiv \dot{a}/a${\ is the Hubble expansion
rate, }$K$ is the 3-curvature constant, and $\rho _{\psi }=\omega \dot{\psi}%
^{2}/2$. For the scalar field, in the absence of non-relativistic matter, we
have 
\begin{equation}
\ddot{\psi}+3{H}\dot{\psi}=0.  \label{fried3}
\end{equation}%
so $\dot{\psi}\propto a^{-3}.$ We should not ignore the possibility that at
high curvatures quantum processes may allow the conversion of $\psi $ energy
into radiation. We take this into account by introducing variable ${\tilde{%
\rho}}_{r}=\rho _{r}e^{-2\psi }$ and rewriting the conservation equations
as: 
\begin{eqnarray}
\dot{\rho}_{\psi }+6H\rho _{\psi } &=&-s({\tilde{\rho}}_{r},\dot{\psi},a),
\label{cons1} \\
\dot{\tilde{\rho}_{r}}+4H{\tilde{\rho}_{r}} &=&s({\tilde{\rho}}_{r},\dot{\psi%
},a).  \label{cons2}
\end{eqnarray}%
In this case, the equation of motion for $\psi $ will contain an additional $%
s$ term which models energy transfer between the $\psi $ field and the
radiation sea in accord with the second law of thermodynamics. We shall
consider the implications of such a process, in all its generality, below,
but let us look first at the equations neglecting this coupling function.

\begin{figure}[tbp]
\centerline{\psfig{file=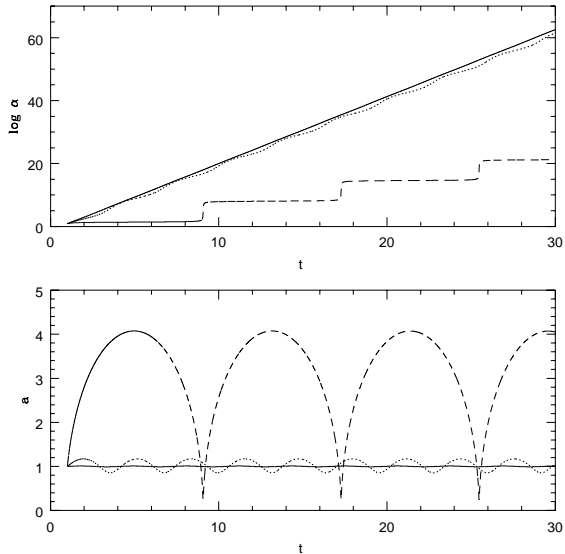,width=8 cm,angle=0}}
\caption{Examples of solutions for $s=0$: a stable, static universe with $%
\protect\alpha $ increasing exponentially (solid line); a stable
perturbation around the static solution (dotted), and a bouncing universe
(dashed). }
\label{fig1}
\end{figure}

\bigskip Consider first a model $s=0$ bouncing universe which is exactly
soluble. Taking $K=+1$ and $\omega <0,$ Eqn. (\ref{fried1}) is 
\begin{equation}
\frac{\dot{a}^{2}}{a^{2}}=-\frac{S}{a^{6}}+\frac{\Gamma }{a^{4}}-\frac{1}{%
a^{2}},  \label{frw}
\end{equation}%
where $S$ and $\Gamma $ are positive constants. In terms of conformal time $%
d\eta =a^{-1}dt$, this can be integrated to give 
\begin{equation}
a^{2}(\eta )=\frac{1}{2}\left[ \Gamma +\sqrt{\Gamma ^{2}-4S}\sin \{2(\eta
+\eta _{0})\}\right]   \label{eta}
\end{equation}%
when $\Gamma ^{2}>4S$. Identifying the expansion maximum and minimum, we see
that $a(\eta )$ is given by 
\begin{equation}
a^{2}=\frac{1}{2}\left[ a_{\max }^{2}+a_{\min }^{2}+(a_{\max }^{2}-a_{\min
}^{2})\sin \{2(\eta +\eta _{0})\}\right] ,  \label{sol2}
\end{equation}%
where $a_{\max }$ is global expansion maximum and $a_{\min }$ is the global
minimum of $a(\eta )$, defined by

\begin{equation}
\begin{array}{c}
a_{\max }^{2} \\ 
a_{\min }^{2}%
\end{array}%
\equiv \frac{\Gamma \pm \sqrt{\Gamma ^{2}-4S}}{2}\   \label{min}
\end{equation}%
Since $\dot{\psi}=Ca^{-3}$ we have, for $\omega <0$, that $S=-\omega C^{2}/2$
and the scalar field driving time-variation of the fine structure constant
is given by

\[
\psi =\pm \frac{2}{\left\vert \omega \right\vert }\tan ^{-1}\left\{ \frac{%
\Gamma \tan (\eta +\eta _{0})+\sqrt{\Gamma ^{2}-4S}}{2\sqrt{S}}\right\} . 
\]%
In Fig.~\ref{fig1} we plot these solutions and show $a(t)$ and $\alpha (t)$
(recall that $\alpha \propto e^{2\psi }$), as functions of \textit{proper}
time, $t$. We note the steady increase of $\alpha $ with time despite the
oscillatory behaviour of the expansion scale factor.

When $s=0$ we have a variety of oscillating solutions whose characteristics
depend on the initial conditions. For bouncing solutions $\alpha $ remains
nearly constant during each cycle but changes sharply, but still
monotonically, at the bounce. There is no significant change of behaviour at
the expansion maximum which also implies that there should be no gross
difference in evolution inside and outside spherical overdensities far from
the bounce. With $a_{max}\gg a_{min}$, and setting $\Gamma ={\tilde{\rho}_{r}%
}a^{4}/3$ and $S=-\rho _{\psi }a^{6}/3$, we have $a_{min}=\sqrt{S/\Gamma }$
and $a_{max}=\sqrt{\Gamma }$. We can then see that the bounce duration is $%
\Delta t\sim a_{min}^{2}/a_{max}$. Since $\dot{\psi}\sim \sqrt{6}\Gamma
^{3/2}/(S\left\vert \omega \right\vert ^{1/2})$ near the bounce, we find
that $\Delta \psi \sim \sqrt{6/\left\vert \omega \right\vert }$,
independently of initial conditions, during each bounce. 

\begin{figure}[tbp]
\centerline{\psfig{file=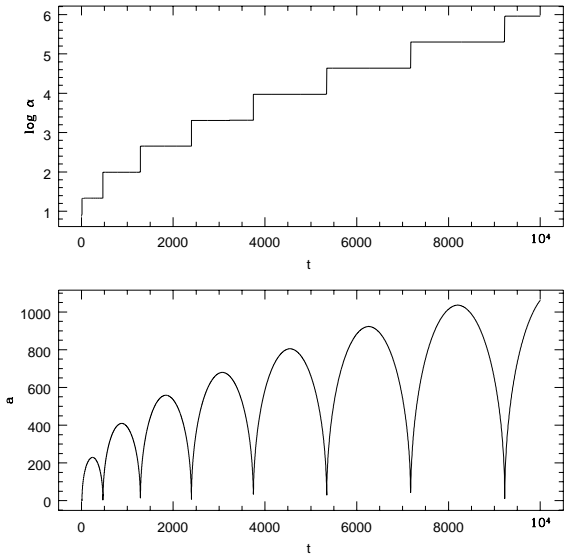,width=8 cm,angle=0}}
\caption{A bouncing universe with $\protect\omega =-0.01$ and $s=100 \protect%
\rho_\protect\protect\psi $. Other couplings lead to qualitatively identical
scenarios. The growth in $a(t)$ and total period from cycle to cycle is
accompanied by increase in $\protect\alpha $ which occurs primarily in the
vicinity of each bounce.}
\label{fig2}
\end{figure}

The extreme case is a stable static universe. Setting $\dot{a}=0$ and $\ddot{%
a}=0$, we can see that this case is realized when $\rho _{\psi }=-{\tilde{%
\rho}}_{r}/2$, giving $a=\sqrt{6/{\tilde{\rho}_{r}}}$. For such a universe $%
\psi $ evolves linearly in $t$, and since
 we have $\alpha \propto e^{2\psi }$ 
there is exponentially rapid increase \cite{bmota}. Even though such a
universe is static, the rulers and clocks of observers change as alpha
changes, so that they actually observe a Milne universe. We can see that the
solution is stable because homogeneous and isotropic perturbations lead to a
universe with regular sinusoidal oscillations as can be seen in Fig.~\ref%
{fig1}. Such solutions are described by (\ref{sol2}) in the case where $%
a_{max}\approx a_{min}$. This situation differs from that found in general
relativity in the absence of ghost fields \cite{btme}.

If $s$ is a non-vanishing then, regardless of its exact functional form,
there are two type of solutions. If $s\neq 0$ at all times, then sooner or
later the universe enters a steady-state evolution with exponential
expansion and constant overall energy density ensured by the appropriate
transfer of energy between the $\psi $ field and radiation. However, we
expect that these energy-transport processes will switch off at low
curvatures, when the universe expands to a sufficiently large size ($%
a>>a_{\min }$) and transport processes become collisionless and far slower
than the expansion rate. Then, the typical evolution is as plotted in Fig.~%
\ref{fig2}. Again, $\psi $ is approximately constant during each cycle and
changes dramatically at the bounce. In addition, each cycle is now bigger
than the previous one, because $\Gamma $ increases at each bounce. 
This is an interesting realisation of the standard Tolman scenario. Cycles
get bigger (and entropy is generated near the bounce) specifically because
radiation is produced from the scalar field close to each bounce. In
producing Fig.~\ref{fig2} we have used $s\propto \rho _{\psi }$ ($\psi $
decays into radiation), but other functional forms may be used with similar
effects.

\begin{figure}[tbp]
\centerline{\psfig{file=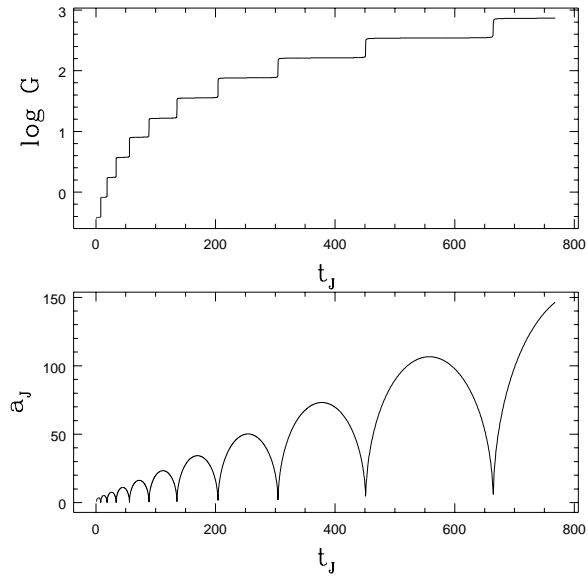,width=8 cm,angle=0}}
\caption{A bouncing universe for a Brans-Dicke theory with $\protect\omega %
_{BD}<-3/2$, in the Jordan frame. In this case the field $\protect\phi $ and
matter are decoupled: they never exchange energy. The amplitude of
oscillations in the scale factor grows monotonically in time and $G$
increases in the Jordan frame. The increases in $G$ occur primarily near the
bounce.}
\label{fig3}
\end{figure}

We now examine similar solutions for the Brans-Dicke (BD) theory of varying $%
G$~\cite{bdicke}. In the Einstein frame, \textit{if the matter content is
pure radiation}, we recover the same equations. All we need to do, then, is
convert the above results into the Jordan frame: the results found for $a$, $%
\rho $, and $\psi $ should then be translated into variables $a_{J}=a/\sqrt{%
\phi }$, $\rho _{J}={\tilde{\rho}}\phi ^{2}$, and $\phi =e^{\psi }$ (the
latter corresponding roughly to $1/G$). Under this transformation we obtain $%
\omega =\omega _{BD}+{\frac{3}{2}}$, where $\omega _{BD}$ in the Brand-Dicke
coupling parameter. Thus we need $\omega _{BD}<-3/2$ for the Brans-Dicke
field to behave like a proper ghost in the Einstein frame. The resulting
dynamics is plotted in Fig.~\ref{fig3}. 

These results may be understood analytically. The essential BD field
equations in the Jordan frame are 
\begin{eqnarray*}
\ddot{\phi}+3H\dot{\phi} &=&0 \\
H^{2} &=&\frac{8\pi \rho }{3\phi }-H\frac{\dot{\phi}}{\phi }+\frac{w_{BD}%
\dot{\phi}^{2}}{6\phi ^{2}}-\frac{K}{a_{J}^{2}}
\end{eqnarray*}%
where overdots now refer to derivatives with respect to $t_{J}$ and $H=\dot{a%
}_{J}/a_{J}$. Hence 
\begin{eqnarray}
\dot{\phi} &=&\frac{A}{a_{J}^{3}}  \label{phi} \\
\frac{\dot{a}_{J}^{2}}{a_{J}^{2}} &=&\frac{\lambda }{a_{J}^{4}\phi }-\frac{%
\dot{a}_{J}}{a_{J}}\frac{\dot{\phi}}{\phi }+\frac{w_{BD}\dot{\phi}^{2}}{%
6\phi ^{2}}-\frac{K}{a_{J}^{2}}  \nonumber
\end{eqnarray}%
with $A$ and $\lambda >0$ constants. We are interested in negative $\omega $
solutions with $K=1$, which give oscillating closed universes (note that $%
\omega _{BD}<0$ is not sufficient for an expansion minimum, we need $\omega
_{BD}<-3/2$). Following the techniques of \cite{jbBD} we put $\frac{A^{2}}{3}%
(2\omega +3)\equiv -C,$ and in the bouncing case $\lambda ^{2}-C>0$, we have
simple exact solution in terms of conformal Jordan time: 
\begin{equation}
\phi a_{J}^{2}=\frac{\lambda }{2}+\frac{1}{2}\sqrt{\lambda ^{2}-C\ }\sin
\{2(\eta +\eta _{0})\}.  \label{y}
\end{equation}%
We see the same behaviour as displayed by the radiation-scalar universe
given above with $s=0$. The minimum value of $\phi a_{J}^{2}$ is $\lambda -%
\sqrt{\lambda ^{2}-C}$ and the maximum is $\lambda +\sqrt{\lambda ^{2}-C}.$
So in conformal time $\phi a_{J}^{2}\approx a_{J}^{2}/G$ undergoes
oscillations of increasing amplitude as the entropy increases (ie if $%
\lambda $ increases in value to model increasing radiation density); that
is, the horizon area ('entropy') in Planck units increases from cycle to
cycle. The full solution for $a(\eta )$ and $\phi (\eta )$ is then obtained
using eqns. (\ref{phi}) and (\ref{y}): 
\begin{equation}
\phi =\phi _{1}\exp \left[ \frac{2A}{\sqrt{C}}\arctan \left[ \frac{\lambda
\tan \{\eta +\eta _{0}\}+\sqrt{\lambda ^{2}-C}}{\sqrt{C}}\right] \right] ,
\label{phisol}
\end{equation}%
with $\phi _{1}$ constant and 
\begin{eqnarray}
a_{J}^{2} &=&\ \phi _{1}^{-1}\left[ \frac{\lambda }{2}+\frac{1}{2}\sqrt{%
\lambda ^{2}-C\ }\sin \{2(\eta +\eta _{0})\}\right] \times   \nonumber \\
&&\exp \left[ -\frac{2A}{\sqrt{C}}\arctan \left[ \frac{\lambda \tan \{\eta
+\eta _{0}\}+\sqrt{\lambda ^{2}-C}}{\sqrt{C}}\right] \right] .  \label{a}
\end{eqnarray}%
The effect of increasing radiation entropy can be seen by increasing the
constant $\lambda $ in these expressions. In this model the field $\phi $
and radiation are fully decoupled. Although the size of the universe
increases in each successive cycle, its size with respect to Planck units
remains the same, unless of course we consider a model in which the field $%
\phi $ and radiation may exchange energy.

We found similar solutions in the BSBM and BD theories because the dynamics
in the Einstein frame is very similar in the absence of non-relativistic
matter. Likewise, one can find identical solutions for the covariant varying
speed of light (VSL) theories described in Refs. \cite{covvsl,rev}. This does
not imply that the VSL, Brans-Dicke and BSBM theories are equivalent; merely
that one needs to add more general matter (even if only as \textquotedblleft
test matter\textquotedblright\ in a radiation-dominated universe) for their
differences to become obvious. Specifically, the coupling to $\mathcal{L}%
_{em}$ in (\ref{Sbsm}) is replaced in Brans-Dicke theory by non-minimally
coupling the matter fields to the metric.

In summary, we have considered some simple exactly soluble models for closed
bouncing universes in theories with varying $\alpha $ and varying $G$ and
examined the effects of simple non-equilibrium behaviour. Even though the
expansion scale factor of the universes undergoes periodic oscillations
about a finite non-singular expansion minimum, we find steady monotonic
change in the values of $\alpha $ and $G$ from cycle to cycle both in the
presence and absence of non-equilibrium behaviour. In the non-equilibrium
case the oscillations of the expansion scale factor grow monotonically in
amplitude and period in accord with the Second Law, and the expansion
dynamics approaches flatness.

\textit{Acknowledgements} We would like to thank B. Carr and C. Hull for
helpful comments.

\end{document}